\documentclass[11pt]{article}
\usepackage{mainsty}
\usepackage{natbib}
\usepackage{enumitem}
\usepackage{diagbox}
\usepackage{xcolor}
\definecolor{ed}{RGB}{225,0,0}

\newcommand{\DIM}{\textsf{DIM}}

\newcommand{\cHT}{\textsf{cHT}}

\newcommand{\obs}{\text{obs}}
\newcommand{\new}{\text{new}}

\title{On the Validity of Fisher Randomization Tests under the Weak Null}

\author{Timothy Sudijono, Edgar Dobriban, Eric Tchetgen Tchetgen}
\date{\today}

\begin{document}
\maketitle

\begin{abstract}
Consider the problem of testing the Neyman weak null of no sample average treatment effect $H_{0N}:\tau = 0$ in randomized experiments with binary outcomes. Although the Fisher Randomization Test (FRT) is typically justified under Fisher's sharp null of no individual treatment effect, we show that it is finite-sample-valid for the weak null $H_{0N}$ in three common combinations of study design and test statistic: 1) a balanced completely randomized experiment with the difference-in-means test statistic, 2) a matched pairs experiment with the same test statistic, and 3) balanced Bernoulli randomized experiments with the centered Horvitz-Thompson test statistic. As an immediate corollary, Fisher's exact test in a balanced completely randomized experiment is valid under the weak null, and so is McNemar's test in a matched pairs experiment. Similar results hold for the one-sided null $H_0:\tau \leq 0$. Beyond balanced designs, we show that twice the one-sided FRT $p$-value with the same test statistics is valid under $H_{0N}$. In simulation experiments, the FRTs compare favorably to other exact procedures.
\end{abstract}

\medskip

\section{Introduction}
\label{sec:introduction}

The Fisher randomization test (FRT) has been used to analyze the results of experiments for almost a century \cite{fisher1930statistical, fisher1935design, pitman1937significance}. The development of the FRT played a crucial role in establishing random allocation of treatments as a fundamental principle in experimental design. Using solely the randomness of this allocation, without other assumptions on the experimental population, the FRT provides a basis for inference on the effect of treatment. For a historical overview and philosophical discussion of the test, see for instance \cite{david2008beginnings,ding2017paradox,  zhang2023randomization, rosenberger202641st}.

The potential outcomes framework \cite{neyman1923application} gives an elegant framework to reason about the test. Consider an experiment with $n$ units, where unit $i$'s potential outcomes under treatment and control conditions are given by $Y_i(1)$ and $Y_i(0)$ respectively, fixed and unknown a priori. The sample average treatment effect $\tau = \frac{1}{n}\sum_{i=1}^n (Y_i(1) - Y_i(0))$ is the parameter of interest. 
Upon administering a randomized treatment $A_i$ to unit $i$ -- where $A_i=1$ denotes treatment and $A_i=0$ denotes control -- and under SUTVA \cite{rubin1980randomization}, we observe only the outcome corresponding to the treatment, 
i.e., $Y_i =Y_i(A_i)$, or equivalently,
$Y_i = A_iY_i(1) + (1-A_i)Y_i(0)$. 
The FRT proceeds by imputing unknown potential outcomes with the observed data assuming the sharp null hypothesis, rerandomizing treatment to obtain a reference distribution, and comparing the observed data to the reference distribution. A precise definition is given in Section \ref{sec:main_results}. As delineated by Welch \cite{welch1937z}, the
usual justification of the FRT comes
from Fisher's sharp null where no unit responds to the treatment:
\begin{equation}
\label{eq:Fisher}
H_{0F}: Y_i(1) = Y_i(0), i=1,\dots,n.
\end{equation}
On the other hand, the scientific question underlying the experiment often involves the Neyman null or \textit{weak null}
representing efficacy on average over the $n$ units:
\begin{align}
\label{eq:Neyman}
& H_{0N}: \tau = 0, \qquad H_{\leq0N}: \tau \leq 0.
\end{align}
The gap between these hypotheses is substantive. 
In many settings, we expect at least some units to respond to the treatment even in a null case, due to inevitable heterogeneity among units. 
Finding examples or classes of FRTs valid under $H_{0N},H_{\leq0N}$ is therefore important. To date, this property has been established in a few settings. One line of work establishes validity under $H_{0N}$ asymptotically \cite{ding2017paradox,wu2021randomization,cohen2022gaussian}, often by considering studentized or pre-pivoted statistics. Caughey et al. \cite{caughey2023randomisation} show the validity of the FRT for bounded null hypotheses of the form $H_0:Y_i(1) - Y_i(0) \leq \delta_i$, for any prespecified constants $\delta_i \in \bR$. More generally, as Ding argues \cite{ding2025randomization}, few examples of tests which are finite-sample valid under $H_{0N}$ are known.
The most notable test is due to Chiba and Rigdon-Hudgens \cite{chiba2015exact,chiba2017stratified,rigdon2015randomization}, which takes a maximum of FRT $p$-values for all potential outcome tables in $H_{0N}, H_{\leq0N}$. Concentration inequalities also enable valid tests \cite{freidling2026randomization, sandoval2026nonasymptotic}, although these are often conservative and can suffer from low power.

This note advances the literature on FRTs and randomization inference by showing that the FRT in three common settings is exactly finite-sample-valid for the weak null. For balanced completely randomized experiments, we show that the one- and two-sided FRTs with the standard difference-in-means statistic are valid for $H_{\leq0N},H_{0N}$, respectively, for all even $n$. For balanced Bernoulli randomized experiments, the FRT with the centered Horvitz-Thompson statistic \cite{aronow2013class,aronow2026minimax, hull2026optimal,sudijono2026sharp} is also valid in finite samples under $H_{0N},H_{\leq0N}$. The same guarantees are true of the FRT in a matched pairs experiment with difference-in-means, where treatment is allocated with a fair coin in every pair of units. As a byproduct, these results establish the weak null validity of two very classical tests: Fisher's exact test in a completely randomized experiment (CRE) and McNemar's test in a matched pairs experiment. Relative to other exact randomization inference procedures \cite{rigdon2015randomization, chiba2017stratified}, running a single FRT may also be computationally more efficient. In the setting of an arbitrary unbalanced or balanced experiment with the estimators as above, we establish a simple rule of thumb that twice the one-sided FRT $p$-value is valid under the weak null. We show in simulation that these three FRTs compare favorably in their respective experimental settings to tests based on concentration inequalities and the Chiba-Rigdon-Hudgens procedure.

\section{Main Results}
\label{sec:main_results}

Before presenting our results, we collect some notation and recall the precise definitions of FRTs. The set of binary potential outcomes in the experiment $\set{(Y_i(1),Y_i(0))}_{i=1}^n$ will be called a \textit{table}. Given a table, define $n_{ab} = \#\set{i:Y_i(1) = a, Y_i(0) = b}$ for $a,b \in \set{0,1}$ to be the sizes of the principal strata \cite{frangakis2002principal}. Given a treatment vector $A^\obs$ drawn from some law $\scr{A}$ called the design, we observe data 
$Y_i(A_i^\obs) = A_i^{\obs} Y_i(1) + (1-A_i^{\obs})Y_i(0)$, $i=1,\dots,n.$ We will refer to data points $Y_i = 1$ as \textit{events} or \textit{ones}. We consider designs where $\Prob(A_i = 1)$, the treatment probability, is constant for all $i$. Call the design \textit{balanced} if the treatment probability is $1/2$ and unbalanced otherwise. Finally, call the design symmetric if $A \stackrel{(d)}{=} 1_n-A$, where $1_n$ is the $n$-dimensional vector of ones.

Let $\hat\tau:\set{0,1}^n \times \set{0,1}^n \rightarrow \bR$ be any statistic of the treatment vector and data. The Fisher randomization test generates $p$-values for the Fisher null $H_{0F}$ as follows. Given the observed data $Y^\obs := Y(A^\obs)$, impute the potential outcome table by assuming $H_{0F}$, i.e., each unit's unobserved outcome equals its observed outcome.
Sample a new treatment vector $A^{\new} \sim \scr{A}$ and set $\hat \tau^{\new} := \hat \tau(A^\new,Y(A^\obs)), \hat \tau^\obs := \hat \tau(A^\obs,Y^\obs)$. Then the one-sided and two-sided FRT $p$-values are defined as
\begin{align}
p_{F}^{+}(A^\obs) := \Prob_{A^{\new} \sim \scr{A}}(\hat\tau^\new \geq \hat\tau^\obs), \\
p_{F}^{\pm}(A^\obs) := \Prob_{A^{\new} \sim \scr{A}}(|\hat\tau^\new| \geq |\hat\tau^\obs|).
\end{align}

We will often omit the dependence on $A^\obs$. Any $p$-value $p$ we discuss in the paper will be called \textit{valid} for $H_{0N}$ at level-$\alpha$ if $\sup_{H_{0N}} \Prob(p \leq \alpha) \leq \alpha.$ We adopt the same definition for $H_{\leq0N}$. We will also call $p$-values with this property \textit{Neyman valid}, or say that they satisfy Neyman validity. 

In this note, we take the test statistic $\hat \tau$ to be the centered Horvitz-Thompson statistic $\hat\tau_{\cHT}$. For experimental designs with treatment probability $\rho,$ the statistic is defined by
\begin{equation}
\label{eq:cHT}
\hat\tau_{\cHT}(A,Y) = \frac{\sum_{i=1}^n A_i (Y_i - 1/2)}{n\rho} - \frac{\sum_{i=1}^n (1-A_i) (Y_i - 1/2)}{n(1-\rho)}.
\end{equation}
The centered Horvitz-Thompson statistic is a special case of a class of estimators introduced in \cite{aronow2013class}. The estimator has recently been studied in several works \cite{aronow2013class, aronow2026minimax, freidling2026randomization, hull2026optimal, sudijono2026sharp} in both survey sampling and experimental settings. Although a strength of the FRT is that any test statistic may be accommodated, this choice is particularly natural and common: $\hat \tau_{\cHT}$ is equivalent to difference-in-means in complete randomization and matched pairs, one of the most widely-used statistics for these experimental designs.

\paragraph{Completely Randomized Experiments.} 

Our main results will show that the FRT based on $\hat \tau_{\cHT}$ is Neyman valid for balanced complete randomization, Bernoulli randomization, and matched pairs. We begin with the balanced completely randomized experiment (CRE). In a balanced CRE, with $n$ even, $n/2$ units are selected uniformly at random to be treated. Since $\hat \tau_{\cHT}$ is equivalent to the difference in means estimator $\hat \tau_{\DIM}$ in a CRE, we discuss them interchangeably. We have the following result.

\begin{theorem}
\label{thm:FRT_CRE_validity}
With a balanced CRE, $p_{F}^{+}$ is valid under the weak null $H_{\leq0N}$ whenever $0 < \alpha \leq 1/2.$ The two-sided FRT $p$-value $p_{F}^{\pm}$ is valid under $H_{0N}$, for all $\alpha \in (0,1)$. 
\end{theorem}

The FRT with difference-in-means in a CRE is known to be equivalent to Fisher's exact test for $2\times 2$ tables \cite{ding2016potential}, one of the oldest tests in statistics \cite{fisher1930statistical}. It is defined as follows. Let $O^\obs$ be the total number of events observed in the data. Let $O_T^\obs$ be the number of events observed in the treatment group.  Under $H_{0F}$, 
$O_T^\obs$ has a hypergeometric distribution
$O_T^\obs \sim \Hypergeom(n,O^\obs,n/2)$. 
Then Fisher's exact test generates a $p$-value defined by $\Prob(\Hypergeom(n,O^\obs,n/2) \geq O_T^\obs)$. The two-sided case is analogous. Theorem \ref{thm:FRT_CRE_validity} extends this validity to the weak null $H_{0N}$.

Another approach to obtain exact tests for $H_{0N}$ based on randomization inference with CRE and $\hat \tau_{\DIM}$ is provided by the procedures of Chiba and Rigdon-Hudgens \cite{chiba2015exact,chiba2017stratified,rigdon2015randomization}. Let $\Delta \in \set{-1,0,1}^n$ be a vector of individual treatment effects compatible with the observed data, satisfying $\sum_i \Delta_i = 0.$ Let $\cl{D}$ be this set of vectors.
Let $p_{\Delta}^\pm$ be the two-sided FRT $p$-value assuming the null $H_{0F,\Delta}: Y_i(1) - Y_i(0) = \Delta_i, i=1,\dots,n$. Taking the worst-case over $\cl{D}$ yields a $p$-value which is valid for $H_{0N}$:
\begin{equation}
\label{eq:rigdon_hudgens_pvalue}
p_{\text{CRH}}^\pm := \sup_{\Delta \in \cl{D}} p_{\Delta}^\pm.
\end{equation}
Throughout this note, we will refer to this test as the Chiba-Rigdon-Hudgens procedure. A similar procedure can be obtained for $H_{\leq0N}$ and with the one-sided FRTs. These approaches require the computation of multiple FRTs, which can be prohibitive in some cases. For the balanced CRE, significant progress has been made. Algorithmic advances \cite{li2016exact,AronowChangLopatto2025} show $O(n\log n)$ FRTs suffice to compute an exact confidence interval for $\tau$. Specializing to testing a single null value $\tau = 0$, the results of \cite{AronowChangLopatto2025} show that $O(n)$ many FRT procedures suffice. Theorem \ref{thm:FRT_CRE_validity} shows that, in fact, 
one FRT is enough to obtain a valid $p$-value. 
Moreover, the two-sided $p$-value $p_{F}^{\pm}$ is at least as powerful as \eqref{eq:rigdon_hudgens_pvalue} by construction and strictly more powerful for some alternatives, as simulation results show in Figure \ref{fig:cre}.

\paragraph{Bernoulli Randomized Experiments.} Next, we consider balanced Bernoulli randomized experiments (BRE) in which treatments are administered via fair coin flips: $A_i,i=1,\dots,n$ are independent Bernoulli random variables with success probability 1/2. Due to operational constraints, BRE appears frequently as the default design in A/B testing and online experimentation \cite{pouget2019testing, kohavi2020trustworthy}. Our second main result shows the validity of the FRT with this test statistic for Bernoulli randomized experiments. 

\begin{theorem}
\label{thm:FRT_BRE_validity}
Under a BRE, the FRT $p$-value $p_{F}^{+}$ constructed using the centered Horvitz-Thompson statistic is valid for the weak null $H_{\leq0N}$ whenever $0 < \alpha \leq 1/2$. For every $\alpha \in (0,1)$, $p_{F}^{\pm}$ is valid for $H_{0N}.$
\end{theorem}

Randomization inference for the BRE appears to be less developed than for the CRE and for matched pairs designs. The most relevant contributions are work by Branson-Bind for the sharp null \cite{branson2019randomization} and recent work by Zhang \cite{zhang2026fast} on exact confidence intervals for $\tau$. The work by Zhang shows for the Horvitz-Thompson estimator, different than \eqref{eq:cHT}, the Chiba-Rigdon-Hudgens procedure $p_{\text{CRH}}^\pm$ of \eqref{eq:rigdon_hudgens_pvalue} can be computed using at most two FRTs. 



\paragraph{Matched Pairs Experiments.} Lastly, consider the setting of a matched pairs experiment. We consider a pool of $n$ units, which are grouped into pairs $p=1,\dots,n/2$. We will index potential outcomes by $(Y_{pi}(1),Y_{pi}(0)),p=1,\dots,n/2,i=0,1$. In a matched pairs experiment, treatment is allocated independently at random in each pair, with exactly one unit receiving treatment. With $A_p$ drawn i.i.d. $\Ber(1/2)$, we observe data 
\begin{align*}
Y_{p1} = Y_{p1}(A_p), \quad Y_{p0} = Y_{p0}(1-A_p), \qquad p=1,\dots,n/2.
\end{align*}
The SATE $\tau$ is often estimated using the difference-in-means estimator $\hat\tau(A,Y) := \frac{2}{n}\sum_{p=1}^{n/2} (2A_p - 1)(Y_{p1} - Y_{p0}),$ which is equal to the centered Horvitz-Thompson estimator \eqref{eq:cHT}. Our third result shows validity in this case.

\begin{theorem}
\label{thm:FRT_matched_pairs_validity}
With a matched pairs experiment and the difference-in-means statistic, $p_{F}^{+}$ is valid under the weak null $H_{\leq0N}$ whenever $0 < \alpha \leq 1/2.$ The two-sided $p$-value $p_{F}^{\pm}$ is valid under $H_{0N}$, for all $\alpha \in (0,1)$.
\end{theorem}

It turns out that the FRT $p$-values $p_F^+, p_F^\pm$ are equal to the one- and two-sided $p$-values obtained from the exact binomial version of McNemar's test \cite[p.~90]{rosenbaum2017observation}, \cite{ding2017paradox}. McNemar's test \cite{mcnemar1947} is a classical procedure used to analyze paired data. In a matched pairs experiment, call a pair $p$ discordant if $(Y_{p1},Y_{p0})$ is $(1,0)$ or $(0,1)$. 
Let $X$ be the total number of discordant pairs, and among these, let $X_+$ be the number of pairs in which the treated unit is $1$. Under $H_{0F}$, $X_+ \mid X \sim \Bin(X,1/2)$. The test rejects if $X_+$ is more extreme than the appropriate $\Bin(X,1/2)$ quantiles, so it is valid conditional on $X$. Theorem \ref{thm:FRT_matched_pairs_validity} shows Neyman validity, unconditionally, of McNemar's test. This appears to be a novel conclusion. 

Several recent works consider exact inference for the matched pairs experiment. Li, Spertus, and Stark \cite{li2025exact} develop the analogs of two confidence interval procedures from \cite{rigdon2015randomization} in general stratified experiments. This includes the inversion of compatible randomization test $p$-values as in \eqref{eq:rigdon_hudgens_pvalue} as well as  Bonferroni combinations of hypergeometric confidence intervals. For matched pairs designs, Zhang \cite{zhang2026fast} shows that computing Rigdon-Hudgens confidence intervals for the matched pairs design has time complexity $O(n \log^2 n)$. Specializing Zhang's results to testing, \eqref{eq:rigdon_hudgens_pvalue} can be computed with at most two FRTs. 
By design, $p_{F}^\pm, p_F^+$ are at least as powerful as $p_{\text{CRH}}^\pm, p_{\text{CRH}}^+$. \cite{wilson2026randomization} develops the analog of the Rigdon-Hudgens attributable effects method by giving analytic solutions to the worst cases in the component randomization tests. Finally, Fogarty \cite{fogarty2023testing} also develops sensitivity analysis procedures for the weak null in matched designs.

\paragraph{Unbalanced Experiments.}

For complete and Bernoulli randomization, it is fairly common to run unbalanced experiments where the treatment probability is $\rho \neq 1/2$. The proof techniques of Theorems \ref{thm:FRT_CRE_validity}, \ref{thm:FRT_BRE_validity} carefully leverage the symmetry of the balanced case and do not extend naturally to the unbalanced case. Instead, we show that a simple, practically appealing fix works: multiply the FRT $p$-value by 2. 

\begin{theorem}
\label{thm:unbalanced_experiments}
Suppose $A^\obs \sim \scr{A}$ is exchangeable. Then $\sup_{H_{\leq0N}} \Prob(p_F^+ \leq \alpha) \leq 2\alpha.$
\end{theorem}

By the same argument, with $p_F^-$ denoting the analogous lower-tail FRT $p$-value, $4\min(p_F^+,p_F^-)$ is a valid $p$-value for $H_{0N}$.
The proof relies
on
making a perhaps unexpected connection between the analysis of FRTs and tournaments, analogously to the analysis of the Jackknife+ \cite{barber2021predictive}.
Since only the exchangeability of the experimental design is used, we expect that the constant $2$ in the guarantee can be sharpened to a constant $C_\rho$ depending on the treatment probability $\rho$. Our previous results show $C_{\rho} \leq 1, \rho = 1/2$, for their respective designs. We leave a sharp characterization of $C_\rho$ to future work.

\section{Simulation Results} For each experimental design, we compare the two-sided FRTs to other exact tests for $H_{0N}$ when $\alpha = 0.05.$ One important class consists of tests based on concentration inequalities, developed in recent work by Freidling and Sandoval et al. \cite{freidling2026randomization, sandoval2026nonasymptotic}, which covers $\hat \tau_{\cHT}$ and every design in this note. Broadly, the results demonstrate that the FRTs are more powerful. For the CRE, we compare $p_{F}^\pm$ to the Chiba-Rigdon-Hudgens $p$-value $p_{\text{CRH}}^\pm$ using the computational speedups from \cite{AronowChangLopatto2025} along with the Barber-Hoeffding and empirical Bennett intervals from \cite[Prop. 4.1, 4.4]{freidling2026randomization}. For the BRE, we compare to $p_{\text{CRH}}^\pm$ with the Horvitz-Thompson estimator, using Zhang's speedups \cite{zhang2026fast}. We also use tests based on the sub-Bernoulli and empirical Bennett intervals for the centered Horvitz-Thompson estimator from \cite[Prop. 3.5, 3.6]{freidling2026randomization}, which builds on \cite{sandoval2026nonasymptotic}.

For each design, we consider three different simulation suites. The first is modeled after clinical trials for rare outcomes. We assume the treatment effect $\tau$ is small, $0.05$, and the control baseline $p_0$---fraction of units with $Y_i(0) = 1$---is small, contained in $\set{0\%,2.5\%,5\%}.$ The fraction of units harmed by the treatment is half of $p_0$. The second suite considers a situation where control events are more common; $p_0$ is 30\%, $\tau$ varies in $\set{0.1,0.2,0.3}$, and the fraction of units harmed is 20\%. Finally, the third suite mimics the second suite, except we impose a monotonicity constraint that there are no units harmed. In each of these suites, we take $n$ to be multiples of $20$ up to $300.$ If $h$ is the fraction of units harmed by the treatment, then the sizes of the principal strata $(n_{11},n_{10},n_{01},n_{00})$ may be expressed in terms of $p_0,h,\tau$ as $n(p_0 - h,\tau + h, h, 1 - p_0 - \tau - h)$.

\begin{figure}[tb]
    \centering
    \includegraphics[width=0.95\linewidth]{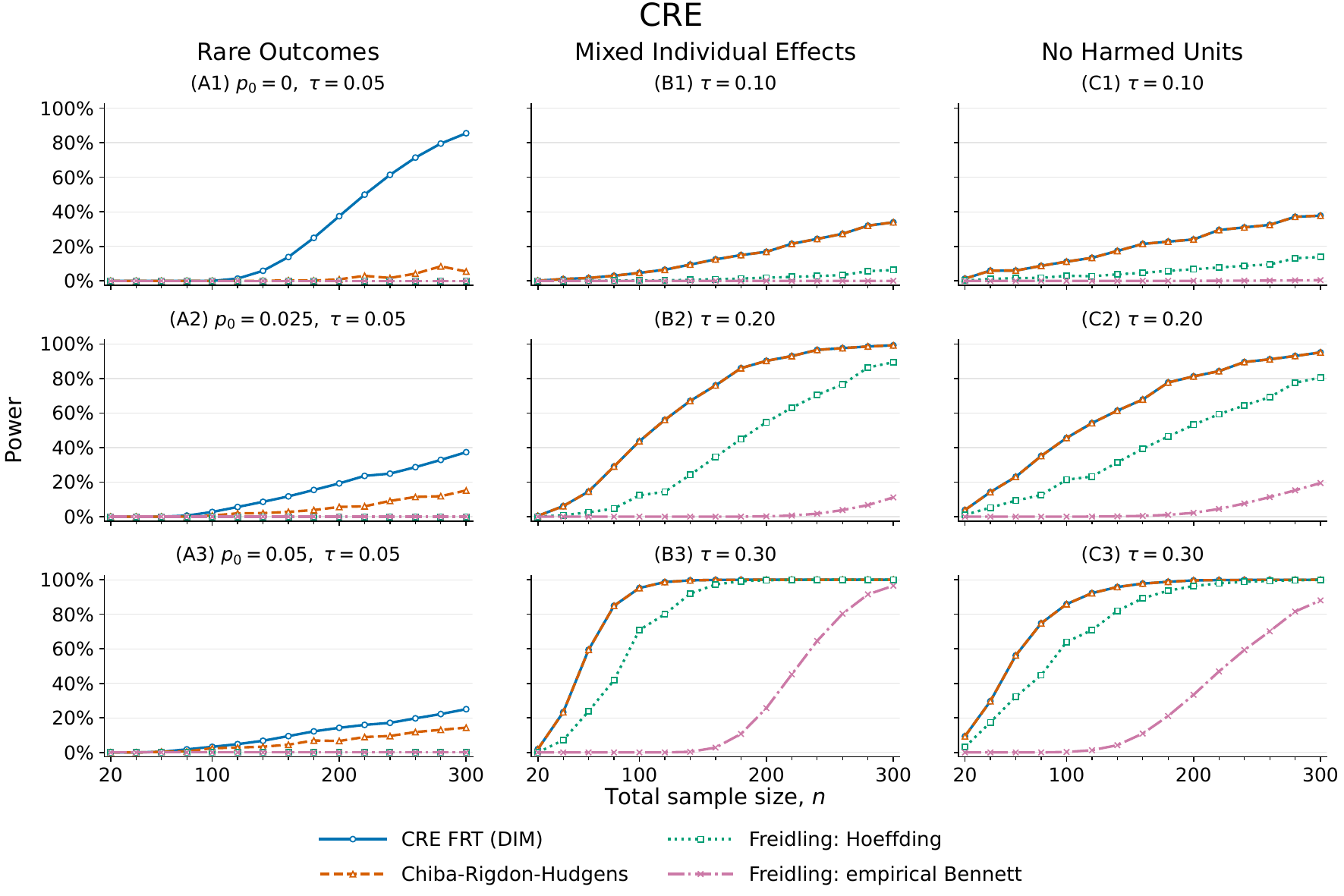}
    \caption{Results on CRE for two-sided tests when $\alpha = 0.05$. Power is computed using 200,000 reassignments of the CRE; $p_0$ denotes the fraction of units with $Y_i(0) = 1$ and $\tau$ is the sample average treatment effect. $p_{\text{CRH}}^\pm$ is computed using the speedups of \cite{AronowChangLopatto2025}.}
    \label{fig:cre}
\end{figure}

\begin{figure}[tb]
    \centering
    \includegraphics[width=0.95\linewidth]{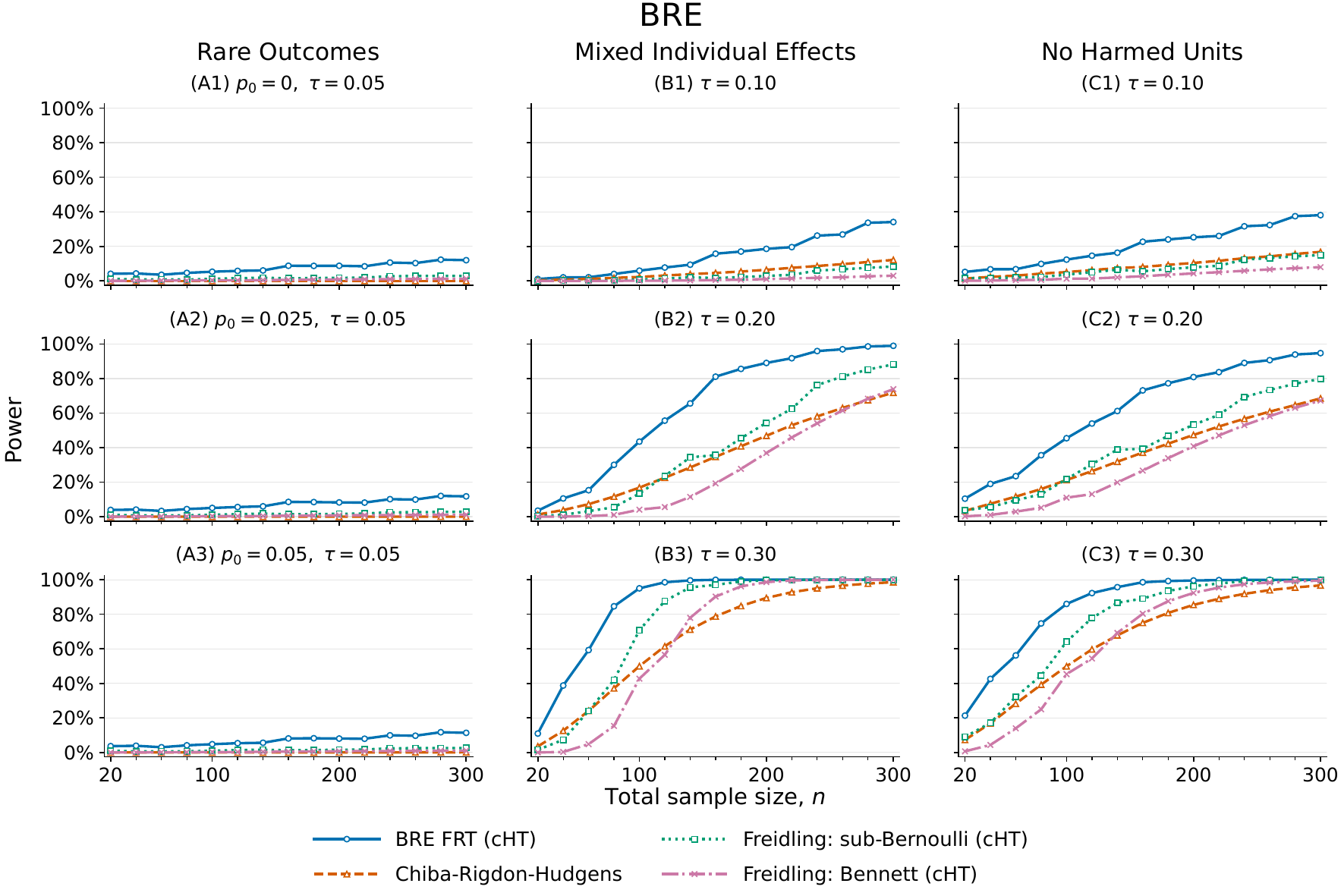}
    \caption{Results for BRE for two-sided tests when $\alpha = 0.05$. Power is computed using 200,000 reassignments of the BRE; $p_0$ denotes the fraction of units with $Y_i(0) = 1$ and $\tau$ is the sample average treatment effect. The $p$-values $p_{\text{CRH}}^\pm$ values are computed using the speedups of Zhang \cite{zhang2026fast}.}
    \label{fig:bre}
\end{figure}

Figures \ref{fig:cre} and \ref{fig:bre} show the power of the comparator procedures in each of these simulation settings, with $\alpha = 0.05$. As all methods in the 3 experimental designs are exactly finite-sample-valid, we do not plot the Type I error. Broadly, the FRT methods outperform the other methods for exact inference, with the concentration inequality based tests being fairly conservative in this binary setting. For the CRE, $p_{\text{CRH}}^\pm$ is indeed different from the FRT $p$-value as can be seen in the first column of Figure \ref{fig:cre}, which highlights a stark gap between the two methods in the rare outcome setting. Outside of this case the two $p$-values are quite similar. Interestingly, the performance of the FRT in panel (A1) is quite promising, suggesting some applications to trials with rare binary outcomes. In the BRE, the FRT with $\hat \tau_{\cHT}$ continues to outperform the concentration inequality based tests along with the exact method of \cite{zhang2026fast}. The performance for rare outcomes is much more modest than that of the FRT in a CRE with difference in means.

\begin{figure}[tb]
    \centering
    \includegraphics[width=0.95\linewidth]{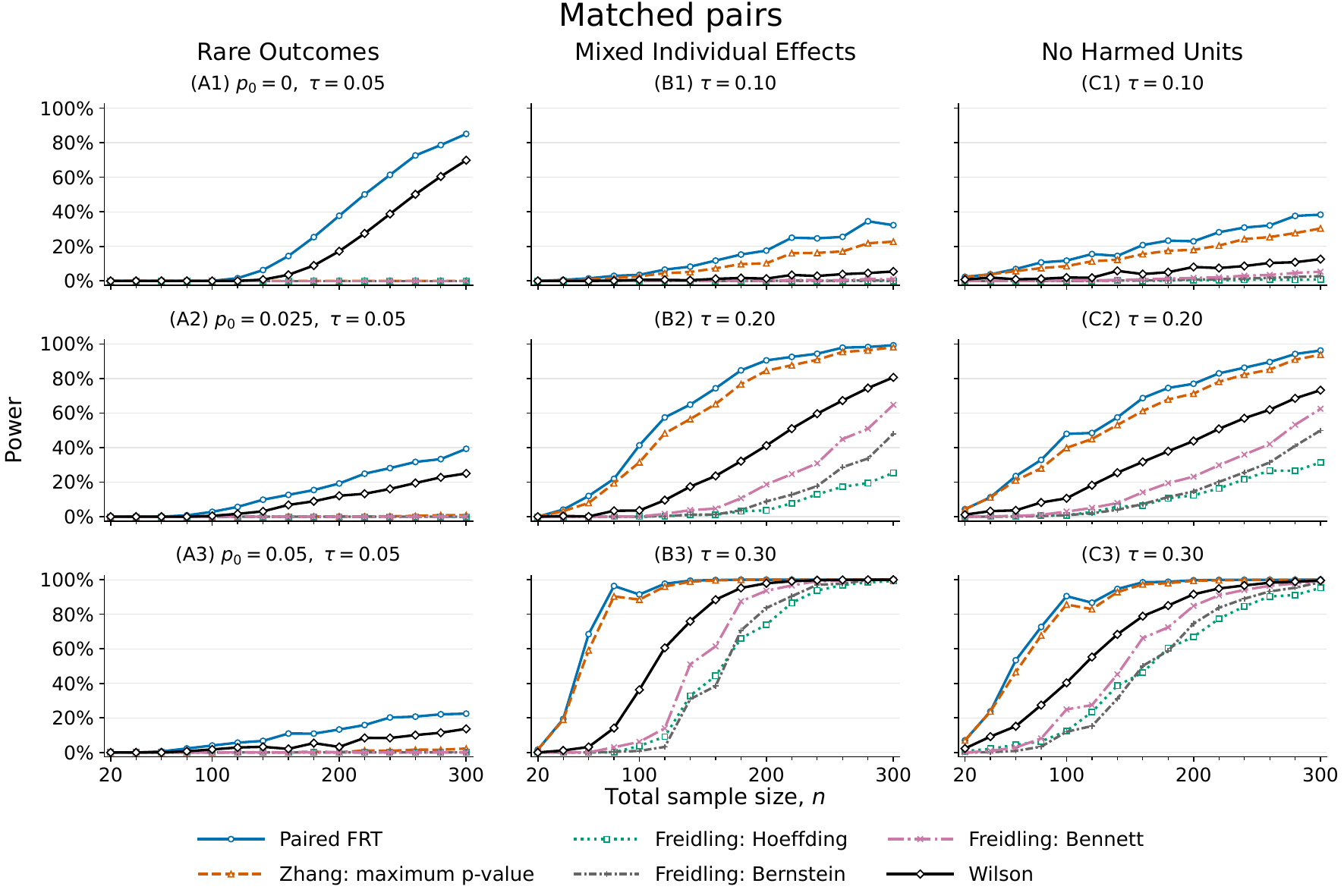}
    \caption{Results for matched pairs experiments for two-sided tests when $\alpha = 0.05$. Power is computed using 200,000 reassignments of the matched pairs design. Matching between units is selected uniformly at random and fixed over the 200,000 runs, for each $n$. Again, $p_0$ denotes the fraction of units with $Y_i(0) = 1$ and $\tau$ is the sample average treatment effect.}
    \label{fig:mp}
\end{figure}

For the matched pairs experiment, we consider the same three simulation suites and use a uniformly chosen pairing between units for each $n$, which is fixed across all reassignments of the design. We compare with the exact RH-type procedure of \cite{zhang2026fast} and tests using the matched pair concentration inequalities in \cite[Prop. 3.6, 3.7]{freidling2026randomization}. Finally, we also compare to the method of Wilson \cite{wilson2026randomization}. The results are pictured in Figure \ref{fig:mp}. The paired FRT behaves similarly to $p_{\text{CRH}}^\pm$, except in the rare outcomes setting where the paired FRT is much more powerful. Wilson's method is also quite competitive, losing only to the paired FRT in the rare outcomes setting and beating the concentration-based procedures in all settings. 

\begin{figure}[tb]
    \centering
    \includegraphics[width=0.95\linewidth]{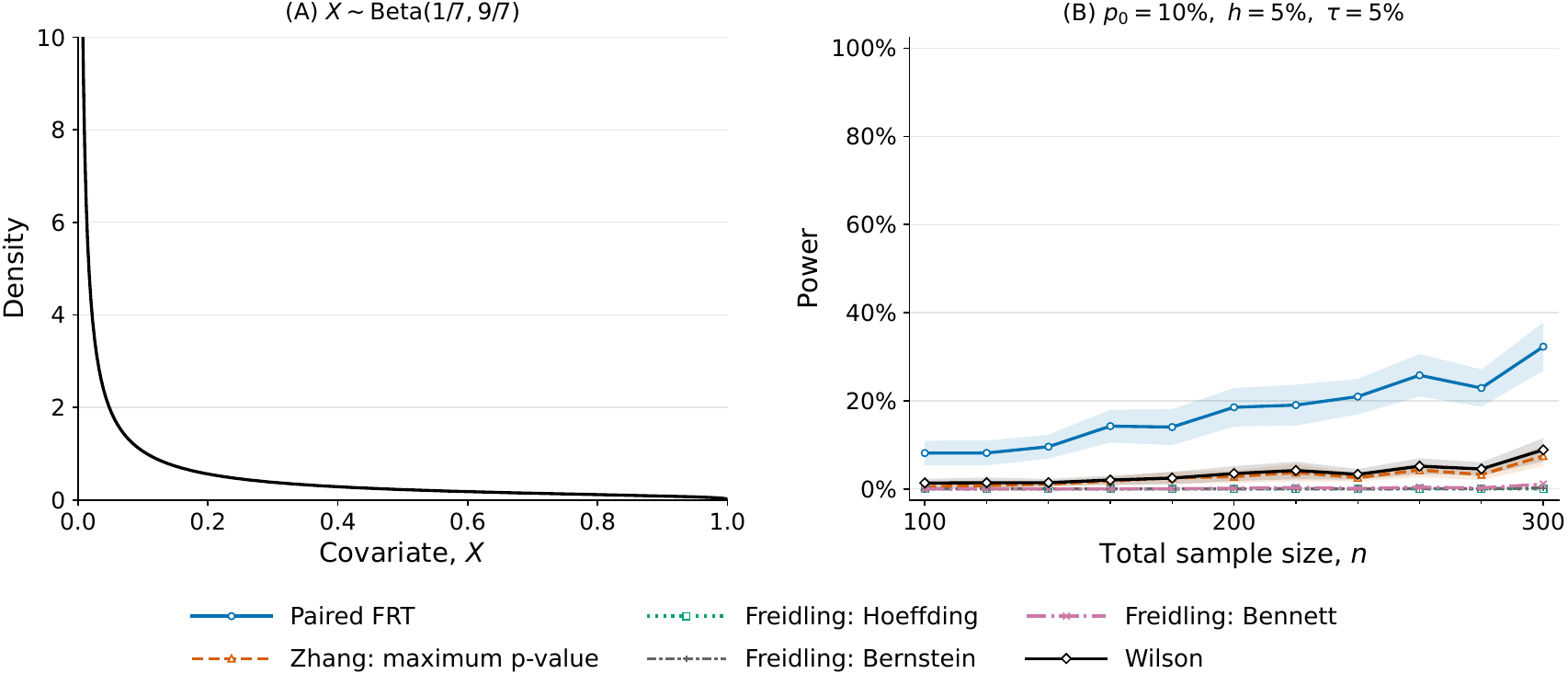}
    \caption{Results for matched pairs experiments, using two-sided tests, $\alpha = 0.05$. Pairs are obtained by matching on an informative covariate $X_i$ generated iid from $\text{Beta}(1/7,9/7)$. Potential outcomes are generated from $X_i$ as described in the text. The left panel shows the density of $X$, and the right panel shows the power of each method, averaged over 100 Monte Carlo redraws of the superpopulation. 95\% pointwise confidence bands are displayed for the variation due to superpopulation draws.}
    \label{fig:mp_informative_covariate}
\end{figure}

Matched pairs designs are often motivated on the basis of blocking on observed information to reduce variation. In Figure \ref{fig:mp_informative_covariate}, we detail an additional simulation setting where the pairs are determined by matching on a scalar covariate. For each unit in the experiment, we sample $X_i \sim \text{Beta}(1/7,9/7)$ and generate $Y_i(0) \mid X_i = x \sim \Ber(x)$. Independently conditional on $X_i$, $Y_i(1) \mid X_i = x$ is generated as $\Ber((1+17x)/18)$. These constants are chosen so that the resulting superpopulation mimics the rare outcome simulation suite described previously, with $p_0 = 0.1, \tau = 0.05$ and the fraction of units harmed being $0.05$. For each $n=100,120,\dots,300$, we sample $X_i$ from the Beta distribution, sort units by $X_i$, and pair adjacent ones. The potential outcomes are generated as above and $(Y_i(1),Y_i(0),X_i)_{i=1}^n$ are held fixed when reassigning the treatment indicators in the matched pairs design. The right panel of Figure \ref{fig:mp_informative_covariate} shows the unconditional power of the previously compared methods, averaged over 100 samples from the superpopulation. The FRT remains the most powerful test; the CRH procedure and Wilson's attributable effects procedure behave similarly.

\section{Discussion}

In this note, we showcase three common circumstances in which the Fisher randomization test is valid under the weak null, allowing users of these tests to significantly upgrade their inferential conclusions. Generalizing and unifying the proof techniques in these theorems may give insights into general conditions for exact finite-sample validity of FRTs. Proving Neyman validity beyond the binary case remains a challenge. In Example \ref{ex:anti_conservative}, we show that the FRT can be anticonservative when the potential outcomes are non-binary. We raise the problem of finding powerful and tractable FRTs, or randomization tests more generally, for bounded outcomes as an important direction for future work.

\paragraph{Acknowledgments.} AI assistance (GPT6) was used in the preparation of this work, including generating figures, suggesting arguments, checking proofs, and revising the article. All proofs were verified by the authors and we take responsibility for any errors.

\bibliographystyle{alpha}
\bibliography{main}

\appendix

\section{Proofs}

In all of the proofs, we will freely use $\e,\e'$ and $B,B'$ to denote independent copies of Rademacher and Bernoulli$(1/2)$ random variables. Let $S_k$ be the simple random walk at time $k$, $\sum_{i=1}^k \e_i$  with $\e_i$ iid Rademacher random variables. For brevity, we will write $\hat \tau_{\cHT}$ simply as $\hat \tau$.

At a technical level, the approach for each design will be to show that the sampling distribution of the statistic---the law of $\hat \tau^\obs$---is more concentrated than the reference distribution---the law of $\hat \tau(A^\new, Y(A^\obs))$---for all realizations of $A^\obs$. At a heuristic level, we should expect this to be true for the centered Horvitz-Thompson estimator on the following grounds. For a balanced design, a brief algebraic manipulation shows that $\hat \tau_{\cHT}$ may be rewritten as
\begin{equation}
\label{eq:alternative_representation_cHT}
\hat \tau_{\cHT}(A,Y) = \tau + \frac{1}{n}\sum_{i=1}^n (2A_i - 1)(Y_i(1) + Y_i(0) - 1).
\end{equation}
Under $H_{0N}$ and a symmetric design, $\hat \tau_{\cHT}$ is symmetrically distributed around zero. 
In particular, $\tau = 0$ and the summands $(2A_i - 1)(Y_i(1) + Y_i(0) - 1)$ are symmetric around zero, taking values in $\set{-1,0,1}$. Notice that $Y_i(1) + Y_i(0) - 1 = 0$ whenever $(Y_i(1),Y_i(0))$ is $(1,0)$ or $(0,1)$. Under $H_{0F}$, each of the summands is nonzero; under any potential outcome table in $H_{0N}$, some of these summands may be zero. Therefore, we expect that the tails of the reference distribution should dominate those of the sampling distribution. Proposition \ref{prop:tail_dominance} shows that it suffices to check this tail dominance conditional on a well-chosen statistic $C$.

\begin{proposition}
\label{prop:tail_dominance}
Consider any symmetric design $A \sim \scr{A}$. For each table in $H_{0N}$, suppose there exists a random variable $C$ such that for all $t > 0$,
\begin{equation}
\label{eq:tail_dominance_criterion}
\Prob_{A^\new}\left(\hat \tau(A^\new,Y(A^\obs)) \geq t\right) \geq \Prob\left(\hat \tau(A^\obs,Y(A^\obs)) \geq t \mid C \right).
\end{equation}
Then, for all $0 < \alpha \leq 1/2$, $\sup_{H_{\leq 0N}}\Prob(p_{F}^+ \leq \alpha) \leq \alpha$. If further $\hat \tau^\obs \mid C$ is symmetric around zero, then $\sup_{H_{0N}} \Prob(p_{F}^\pm \leq \alpha) \leq \alpha$.
\end{proposition}

\begin{proof}
We first prove the guarantees for $H_{0N}$. If $A$ is symmetric, then \eqref{eq:alternative_representation_cHT} shows that $\hat \tau(A^\new,Y(A^\obs))$ is symmetric around zero conditional on $A^\obs$. For $\alpha \leq 1/2$, $p_F^+ \leq \alpha$ implies $\hat \tau^\obs > 0$ by this symmetry. Let $\bar{F}_C(t) := \Prob\left(\hat \tau^\obs \geq t \mid C \right)$ be the conditional sampling distribution. By \eqref{eq:tail_dominance_criterion} and the definition of the FRT $p$-value, $\set{p_F^+ \leq \alpha} \subseteq \set{\bar{F}_C(\hat \tau^\obs) \leq \alpha}.$ Then $\Prob(p_F^+ \leq \alpha \mid C) \leq \Prob(\bar{F}_C(\hat \tau^\obs) \leq \alpha \mid C) \leq \alpha$. Take expectations to conclude the validity of $p_F^+.$

By assumption, $\hat \tau^\obs$ is symmetric around zero conditional on $C.$ Multiplying each side of \eqref{eq:tail_dominance_criterion} by two and using this symmetry and that of $\hat \tau^\new \mid A^\obs$, we obtain
\[
\Prob_{A^\new}\left(|\hat \tau(A^\new,Y(A^\obs))| \geq t\right) \geq \Prob\left(|\hat \tau(A^\obs,Y(A^\obs))| \geq t \mid C \right).
\]
The same argument shows the conditional validity of $p_F^\pm$. To extend to $H_{\leq 0N}$, we exploit the following representation of $\hat \tau_{\cHT}$:
\begin{equation}
\hat \tau(A,Y) = \frac{1}{n}\sum_{i=1}^n (2A_i - 1)(2Y_i - 1).
\end{equation}
Then, with $D(A^\new,A^\obs) := \hat \tau(A^\new,Y(A^\obs)) - \hat \tau(A^\obs,Y(A^\obs))$,
\[
D(A^\new,A^\obs) = \frac{2}{n}\sum_{i=1}^n (A_i^\new - A_i^\obs)(2Y_i^\obs - 1),
\]
and $p_F^+(A^\obs) = \Prob_{A^\new}(D(A^\new,A^\obs) \geq 0)$. For any table $\cl{P} = \set{(Y_i(1),Y_i(0))_{i=1}^n}$ in $H_{\leq 0N}$, there are at least $|n\tau|$ units with potential outcomes $(0,1)$. Take one of these units and  switch its potential outcomes to $(1,1)$ to obtain a table $\cl{P}'.$ Then the difference of $D(A^\new,A^\obs)$ under $\cl{P}'$ and $\cl{P}$ is given by $\frac{4}{n}(A_i^\new - A_i^\obs)A_i^\obs \leq 0.$ Thus, $D(A^\new,A^\obs)$ decreases when moving from $\cl{P}$ to $\cl{P}'$, and so $p_F^+(A^\obs)$ decreases. This yields $\Prob_{\cl{P}}(p_F^+ \leq \alpha) \leq \Prob_{\cl{P}'}(p_F^+ \leq \alpha).$ Iterate this guarantee until a table $\cl{P}' \in H_{0N}$ is obtained. Then, use the first part of the proposition.
\end{proof}

Before treating each of the different designs, we record the following useful tail comparison lemma.

\begin{lemma}[Random walk tail comparison]
\label{lemma:random_walk_tail_comparison}
For any nonnegative integers $k' > k$ of the same parity and an integer threshold $t > 0$ of the same parity as $k$, $\Prob(S_{k'} \geq t) \geq \Prob(S_{k} \geq t)$.
\end{lemma}
\begin{proof}
It suffices to show that $\Prob(S_{k+2} \geq t) \geq \Prob(S_{k} \geq t)$ for every $k$ and $t > 0.$ By conditioning on the last two increments of $S_{k+2}$, we have
\[
\Prob(S_{k+2} \geq t) = \frac{1}{4}\Prob(S_{k} \geq t+2) + \frac{1}{2}\Prob(S_{k} \geq t) + \frac{1}{4}\Prob(S_{k} \geq t-2).
\]
Thus, $\Prob(S_{k+2} \geq t) \geq \Prob(S_{k} \geq t)$ is equivalent to $\Prob(S_{k} \geq t+2) - \Prob(S_{k} \geq t) \geq \Prob(S_{k} \geq t) - \Prob(S_{k} \geq t-2)$, or $\Prob(S_k = t) \leq \Prob(S_k = t - 2)$. This is true by unimodality of $S_k$. 
\end{proof}

\subsection{Proofs for CRE}

Throughout this section, we will refer to units $i$ for which $Y_i = 1,A_i = 1$ as \textit{treated ones}, $Y_i = 1,A_i = 0$ as \textit{control ones}, and so forth for the remaining combinations. Under a CRE, let $O_T^\obs = \#\set{i:A_i^\obs = 1, Y_i(A_i^\obs) = 1}, O_C^\obs = \#\set{i:A_i^\obs = 0, Y_i(A_i^\obs) = 1}$ be the number of treated ones and control ones respectively. Let $O^\obs = O_T^\obs + O_C^\obs = \sum_i Y_i$ be the total number of ones in the data. Let $\Hypergeom(N,K,n)$ denote the hypergeometric law with total urn size $N$, number of successes $K$, and number of draws $n$.

Given observed data vector $y \in \set{0,1}^n$ and treatment vector $a \in \set{0,1}^n$ with $o_T$ many treated ones and $o_C$ many control ones, call a table $(Y_i(1),Y_i(0))_{i=1}^n$ \textit{compatible} with $y,a$ if $y_i = Y_i(a_i)$ for all $i$. Let $N(o_T,o_C)$ be the number of potential outcome tables compatible with $y$ that have strata $(n_{11}, n_{10}, n_{01}, n_{00})$. Lemma \ref{lemma:table_completions} shows that $N$ only depends on $y$ through $o_T,o_C$. 

\begin{proof}[Proof of Theorem \ref{thm:FRT_CRE_validity}]

We will apply Proposition \ref{prop:tail_dominance} by checking \eqref{eq:tail_dominance_criterion} with $C = O^\obs$. Under a CRE, $\hat \tau^\obs$ is equal to $\frac{2}{n}(O_T^\obs - O_C^\obs) = \frac{2}{n}(2O_T^\obs - O^\obs)$. It is easy to see that the reference distribution, the law of $\hat \tau(A^\new, Y(A^\obs)) \mid A^\obs$, is equal to the law of
\[
\frac{2}{n}\left(2H - O^\obs \right), \qquad H \sim \Hypergeom(n,O^\obs,n/2),
\]
which only depends on $A^\obs$ through $O^\obs$. We check that 
\begin{equation}
\label{eq:stochastic_dominance}
\Prob \left( \left|H - \frac{O^\obs}{2}\right| \geq t \mid O^\obs \right)  \geq \Prob\left(\left|O^\obs_T - \frac{O^\obs}{2}\right| \geq t \mid O^\obs \right).
\end{equation}
Symmetry of both $H - O^\obs/2 \mid O^\obs$ and $O_T^\obs - O^\obs/2 \mid O^\obs$ shows that the one-sided guarantee holds. To prove \eqref{eq:stochastic_dominance}, we consider the likelihood ratio of the conditional distributions of $O^\obs_T$ and $H$, given $O^\obs$. First, we calculate $\Prob(O_T^\obs = o_T, O_C^\obs = o_C)$. Complete randomization is combinatorially equivalent to uniformly permuting the units and treating the first $n/2$ units in the permuted order. We count the number of permutations for which the observed data consists of $o_T$ many ones among the first $n/2$ units and $o_C$ many ones in the remaining $n/2$ units. There are $\binom{n/2}{o_T}$ many ways to select the locations of the treated ones and $\binom{n/2}{o_C}$ many ways to select the locations of the control ones. There are $N(o_T,o_C)$ many tables with the proper strata that are compatible with the chosen locations. Finally, the units may be permuted within each stratum, giving a factor of $n_{11}!n_{10}!n_{01}!n_{00}!$. Thus,
\begin{align*}
\Prob(O_T^\obs = o_T, O_C^\obs = o_C) = \frac{n_{11}!n_{10}!n_{01}!n_{00}!\binom{n/2}{o_T}\binom{n/2}{o_C} N(o_T,o_C)}{n!}.
\end{align*}
Write $\Prob_o(H = k)$ as shorthand for $\Prob(H = k \mid O^\obs = o).$ By rewriting the hypergeometric mass function, $\Prob_o(H = k) = \frac{\binom{n/2}{k}\binom{n/2}{o-k}}{\binom{n}{o}}.$ The likelihood ratio then satisfies
\begin{equation}
R(k) := \frac{\Prob(O_T^\obs = k \mid O^\obs = o)}{\Prob_o(H = k)} \propto N(k, o- k).
\end{equation}
Proposition \ref{prop:table_completion_count_ordering} shows that for every $o_T,o_C$ also $N(o_T,o_C) = N(o_C,o_T)$ and when $o_T \geq o_C+2, N(o_T,o_C) \leq N(o_T-1,o_C+1)$. Therefore, $N(k,o-k)$ is nonincreasing as $|k - o/2|$ increases. Because $\sum_k R(k) \Prob_o(H = k) = 1$, the likelihood ratio $R(k)$ crosses $1$ at most once, from above to below, as $k$ ranges over $\floor{o/2+1},\dots,o$. Call this crossing point $k^*$, so $R(k) \leq 1, k > k^*$ and $R(k) > 1, k=\floor{o/2+1},\dots,k^*$.

Fix any $t > k^* - o/2$. Then
\begin{align}
\Prob(|O_T^\obs - o/2| \geq t  \mid O^\obs = o) & = \sum_{|k - o/2| \geq t} R(k) \Prob_o(H = k) \leq \sum_{|k - o/2| \geq t} \Prob_o(H = k) = \Prob(|H - o/2| \geq t).
\end{align}
When \(t \leq k^* - o/2\), redo the above derivation with the lower tail event and take complements. This proves \eqref{eq:stochastic_dominance}. Proposition \ref{prop:tail_dominance} gives the one and two-sided conclusions. 
\end{proof}

\begin{lemma}
\label{lemma:table_completions}
The completion count $N(o_T,o_C)$ only depends on $a,y \in \set{0,1}^n$ through $o_T,o_C$.
\end{lemma}
\begin{proof}
Take any two data vectors $y,y' \in \set{0,1}^n$ with the same number of treated ones and control ones $o_T,o_C.$ Take a permutation $\sigma$ for which $y' = \sigma y$ and $a' = \sigma a$. For any table $\cl{P}$ which is compatible with $y$, applying the permutation $\sigma$ to the rows of $\cl{P}$ yields another table $\cl{P}'$ compatible with $y'$. This mapping is a bijection between the set of tables compatible with $y$ and those compatible with $y'$. Therefore the number only depends on $o_T,o_C$.
\end{proof}

\begin{proposition}
\label{prop:table_completion_count_ordering}
Suppose $n_{01} = n_{10}$, $o_C \geq 0$, and $o_C + 2 \leq o_T \leq n/2.$ Then
$N(o_T,o_C) \leq N(o_T - 1, o_C + 1).$ For any $o_C,o_T, N(o_T,o_C) = N(o_C,o_T).$
\end{proposition}
\begin{proof}
Let $\cl{N}(o_T,o_C)$ be the set of tables with principal strata counts $(n_{11},n_{10},n_{01},n_{00})$ and first $n/2$ units treated, such that $Y_i(1) = \one\set{i \leq o_T}$ and $Y_{n/2 + i}(0) = \one\set{i \leq o_C}$ for $i = 1,\dots,n/2$. These are compatible with the observed data $(o_T,o_C)$ and by Lemma \ref{lemma:table_completions}, $|\cl{N}(o_T,o_C)| = N(o_T,o_C).$ We will exhibit an injection $\Phi:\cl{N}(o_T,o_C) \hookrightarrow \cl{N}(o_T-1,o_C+1)$.

To define the injection, fix a table $\cl{P} \in \cl{N}(o_T,o_C)$. See the left table of Figure \ref{fig:row-blocks} for a running visualization. We define five disjoint subsets of its rows. Let $W = \set{o_T,n/2+o_C+1}$ be the row indices of the last treated 1 and the first control 0. Let $V_T$ be the treated zeros $\set{o_T + 1,\dots,n/2}$ and $V_C$ the control ones $\set{n/2+1,\dots,n/2+o_C}.$ Finally, let $U_T$ consist of the first $o_C$ treated ones $\set{1,\dots,o_C}$ and $U_C$ be the first $n/2 - o_T$ control zeros, excluding the row in $W$: $\set{n/2+o_C+2,\dots,n+1+o_C - o_T}$. Because $o_T \geq o_C+2$, $|U_T| = |V_C|$ and $|V_T| = |U_C|$.

The idea is to swap the rows in $W$ and reverse their potential outcomes to define a table $\Phi(\cl{P})$ in $\cl{N}(o_T-1,o_C+1)$. The issue is that the principal strata counts will not be maintained. We compensate for this by further swapping rows between $U_T, V_C$ and $V_T,U_C$. To this end, we define a pairing of row indices by 1) pairing the rows in $W$, 2) pairing the rows in $U_T$ with the corresponding rows in $V_C$, in increasing row order, and 3) similarly with $V_T$ to $U_C$. Define a list of row pairs $\set{(p_j,q_j)}_{j=1}^K, K = 1 + |U_T \cup U_C|$ by concatenating the matched pairs in $W$, then those in $(U_T,V_C)$ followed by $(V_T, U_C)$. Explicitly,
\begin{equation}
(p_k,q_k) = 
\begin{cases}
(o_T,n/2+o_C+1) & \text{ if } k = 1\\
(k-1,n/2 + k - 1) & \text{ if } k = 2,\dots,o_C + 1 \\
(k + o_T - o_C - 1, k + n/2) & \text{ if } k > o_C + 1.
\end{cases}
\end{equation}
Note that this pairing forces 
\begin{equation}
\label{eq:pairing_property}
Y_{p_k}(1) = Y_{q_k}(0), \quad \forall k \geq 2.
\end{equation}

Define
\begin{equation}
    \delta_k(\cl{P}) := \sum_{i=1}^k\left(Y_{p_i}(1) - Y_{p_i}(0) + Y_{q_i}(1) - Y_{q_i}(0) \right), \quad k = 1,\dots,K,
\end{equation}
to be the running imbalance between the pairs in this list. Let $k_*(\cl{P})$ be the first index for which this running sum hits zero. This index always exists. By construction of the set $W$, $\delta_1(\cl{P}) \geq 0$. Because $n_{10} = n_{01}$ and $V_T \cup V_C$ contains all rows with type $(0,1)$, $\delta_K(\cl{P}) \leq 0$. Finally, \eqref{eq:pairing_property} forces $|\delta_k(\cl{P})-\delta_{k-1}(\cl{P})| \leq 1$ for $k \geq 2.$

Define the injection $\Phi$ as follows. For each $1 \leq j \leq k_*(\cl{P})$, swap the rows $p_j, q_j$, then swap the potential outcomes within each row. This results in a table $\Phi(\cl{P})$ with rows $(Y_i'(1),Y_i'(0))_{i=1}^n$ defined by
\begin{equation}
Y'_{p_i}(a) = Y_{q_i}(1-a), \qquad Y'_{q_i}(a) = Y_{p_i}(1-a), \qquad 1 \leq i \leq k_*(\cl{P}), \quad a \in \set{0,1},
\end{equation}
and all other rows unchanged. The swaps exchange types $(1,0)$ and $(0,1)$ while preserving $(1,1)$ and $(0,0)$. Since $\delta_{k_*(\cl{P})}(\cl{P}) = 0$, the numbers of all principal strata are preserved. The observed entries only change for the rows in $W$. Thus $\Phi(\cl{P}) \in \cl{N}(o_T-1,o_C+1)$.

Finally, we show this map is an injection. Consider a table $\cl{P}' \in \cl{N}(o_T - 1, o_C+1)$ that is equal to $\Phi(\cl{P})$ for some $\cl{P}$. Given $\cl{P}'$, flipping the first treated zero to $1$ and last control one to $0$ yields the same observed data pattern as $\cl{P},$ allowing us to recover the same sets $W, U_T,U_C,V_T,V_C$. For all $k \leq k_*$, we have $\delta_k(\cl{P}') = -\delta_k(\cl{P})$. Thus, $k_*(\cl{P}') = k_*(\cl{P})$. Applying the same swap-reverse transformations as in $\Phi$ to $\cl{P}'$ therefore recovers the original table $\cl{P}.$

\textit{Symmetry.} For any table $\cl{P}$ in $\cl{N}(o_T,o_C)$, create a new table by switching each unit's outcomes from $(Y_i(1),Y_i(0))$ to $(Y_i(0), Y_i(1))$ and flipping treatment and control. Because $n_{10} = n_{01}$, this creates another table in $\cl{N}(o_C,o_T)$. This transformation is a bijection. Therefore $N(o_T,o_C) = N(o_C,o_T)$.

\end{proof}

\begin{figure}[H]
\centering
\begin{minipage}[c]{0.44\textwidth}
\centering
$\cl{P}\in\cl{N}(5,3)$
\par\smallskip
\begin{tikzpicture}[x=1.2cm,y=.36cm,font=\small]
\node at (1.15,.75) {Unit};
\node at (1.95,.75) {$Y_i(1)$};
\node at (2.85,.75) {$Y_i(0)$};
\node at (3.85,.75) {Label};
\foreach \i/\a/\b/\block/\shade in {
 1/1/\unobs{1}/U_T/blockone,
 2/1/\unobs{1}/U_T/blockone,
 3/1/\unobs{1}/U_T/blockone,
 4/1/\unobs{1}/\unpairedlabel/unpairedrow,
 5/1/\unobs{0}/W/blockc,
 6/0/\unobs{0}/V_T/blocktwo,
 7/0/\unobs{0}/V_T/blocktwo,
 8/\unobs{0}/1/V_C/blocktwo,
 9/\unobs{0}/1/V_C/blocktwo,
 10/\unobs{1}/1/V_C/blocktwo,
 11/\unobs{1}/0/W/blockc,
 12/\unobs{0}/0/U_C/blockone,
 13/\unobs{0}/0/U_C/blockone,
 14/\unobs{0}/0/\unpairedlabel/unpairedrow}{
 \fill[\shade!10] (1.50,-\i+1) rectangle (4.25,-\i);
 \node[text=black!65] at (1.15,-\i+.5) {\i};
 \node at (1.95,-\i+.5) {$\a$};
 \node at (2.85,-\i+.5) {$\b$};
 \node[text=\shade] at (3.85,-\i+.5) {$\block$};
 \draw[black!18] (1.5,-\i) -- (4.25,-\i);
}
\draw[black!40] (1.5,0) rectangle (3.3,-14);
\draw[black!25] (2.4,0) -- (2.4,-14);
\foreach \top/\bottom in {4/5,10/11}{
 \draw[blockc,thick] (1.5,-\top) rectangle (4.25,-\bottom);
}
\draw[black!65] (.9,-7) -- (4.25,-7);
\draw[black!50] (.75,0) -- (.6,0) -- (.6,-7) -- (.75,-7);
\draw[black!50] (.75,-7) -- (.6,-7) -- (.6,-14) -- (.75,-14);
\node[rotate=90] at (.15,-3.5) {Treated};
\node[rotate=90] at (.15,-10.5) {Control};
\end{tikzpicture}
\end{minipage}\hfill
\begin{minipage}[c]{0.08\textwidth}
\centering
$\xrightarrow{\;\Phi\;}$
\end{minipage}\hfill
\begin{minipage}[c]{0.44\textwidth}
\centering
$\cl{P}'=\Phi(\cl{P})\in\cl{N}(4,4)$
\par\smallskip
\begin{tikzpicture}[x=1.2cm,y=.36cm,font=\small]
\node at (1.15,.75) {Unit};
\node at (1.95,.75) {$Y'_i(1)$};
\node at (2.85,.75) {$Y'_i(0)$};
\node at (3.85,.75) {Label};
\foreach \i/\a/\b/\block/\shade in {
 1/1/\unobs{\mathbf{0}}/U_T/blockone,
 2/1/\unobs{\mathbf{0}}/U_T/blockone,
 3/1/\unobs{1}/U_T/blockone,
 4/1/\unobs{1}/\unpairedlabel/unpairedrow,
 5/\mathbf{0}/\unobs{\mathbf{1}}/W/blockc,
 6/0/\unobs{0}/V_T/blocktwo,
 7/0/\unobs{0}/V_T/blocktwo,
 8/\unobs{\mathbf{1}}/1/V_C/blocktwo,
 9/\unobs{\mathbf{1}}/1/V_C/blocktwo,
 10/\unobs{1}/1/V_C/blocktwo,
 11/\unobs{\mathbf{0}}/\mathbf{1}/W/blockc,
 12/\unobs{0}/0/U_C/blockone,
 13/\unobs{0}/0/U_C/blockone,
 14/\unobs{0}/0/\unpairedlabel/unpairedrow}{
 \fill[\shade!10] (1.50,-\i+1) rectangle (4.25,-\i);
 \node[text=black!65] at (1.15,-\i+.5) {\i};
 \node at (1.95,-\i+.5) {$\a$};
 \node at (2.85,-\i+.5) {$\b$};
 \node[text=\shade] at (3.85,-\i+.5) {$\block$};
 \draw[black!18] (1.5,-\i) -- (4.25,-\i);
}
\draw[black!40] (1.5,0) rectangle (3.3,-14);
\draw[black!25] (2.4,0) -- (2.4,-14);
\foreach \top/\bottom in {4/5,10/11}{
 \draw[blockc,thick] (1.5,-\top) rectangle (4.25,-\bottom);
}
\draw[black!65] (.9,-7) -- (4.25,-7);
\draw[black!50] (.75,0) -- (.6,0) -- (.6,-7) -- (.75,-7);
\draw[black!50] (.75,-7) -- (.6,-7) -- (.6,-14) -- (.75,-14);
\node[rotate=90] at (.15,-3.5) {Treated};
\node[rotate=90] at (.15,-10.5) {Control};
\end{tikzpicture}
\end{minipage}
\caption{Example of the injection $\Phi$ with $(n_{11},n_{10},n_{01},n_{00})=(5,2,2,5)$ and $o_T = 5,o_C = 3$. The ordered pairs are $(5,11)$, $(1,8)$, $(2,9)$, $(3,10)$, $(6,12)$,
and $(7,13)$. The first three pairs are swapped and flipped, with changed entries in bold font. Grey entries are unobserved.}
\label{fig:row-blocks}
\end{figure}

\subsection{Proofs for Bernoulli Randomized Experiments}

\begin{proof}[Proof of Theorem \ref{thm:FRT_BRE_validity}]
We will show that \eqref{eq:tail_dominance_criterion} holds with $C = 0.$
Recall that $S_k$ denotes the simple random walk at time $k$. By \eqref{eq:alternative_representation_cHT}, under $H_{0N}$, $\hat \tau^\obs$ is equal in distribution to $\frac{1}{n}S_{n_{11}+n_{00}}$. Using \eqref{eq:alternative_representation_cHT} again, $\hat{\tau}_{\cHT}(A^{\new},Y(A^\obs)) = \frac{1}{n}\sum_{i=1}^n (2A^{\new}_i - 1)(2Y_i(A_i^\obs) - 1).$ Since $2Y_i(A_i^\obs) - 1 \in \set{-1,1}$, the reference distribution of the FRT is equal to $\frac{1}{n}S_n$, regardless of $A_i^\obs$. Under $H_{0N}, n_{10} = n_{01}$, so $n \equiv n_{11} + n_{00} \mod 2$. Lemma \ref{lemma:random_walk_tail_comparison} shows 
\[
\Prob\left(\frac{1}{n }S_n \geq t \right) \geq \Prob\left(\frac{1}{n }S_{n_{11} + n_{00}} \geq t \right).
\]
Apply Proposition \ref{prop:tail_dominance} to conclude.
\end{proof}

\subsection{Proofs for Matched Pairs Experiments}

\begin{proof}[Proof of Theorem \ref{thm:FRT_matched_pairs_validity}]

In the matched pairs experiment, denote the summands in the difference in means estimator by $\delta_p^\obs := (2A^\obs_p - 1)(Y_{p1}(A_p^\obs) - Y_{p0}(1-A_p^\obs))$, or equivalently,
\begin{equation}
\label{eq:matched_pairs_dim_summand}
\delta_p^\obs = A_p^\obs(Y_{p1}(1) - Y_{p0}(0)) + (1-A^\obs_p)(Y_{p0}(1) - Y_{p1}(0)).
\end{equation}
Let $\delta^{\new}_p$ be the summand in the resampled estimator, which by \eqref{eq:matched_pairs_dim_summand} is equal to $A_p^{\new}(Y_{p1}(A_p^\obs) - Y_{p0}(1-A_p^\obs)) + (1-A^{\new}_p)(Y_{p0}(1-A_p^\obs) - Y_{p1}(A_p^\obs))$. Thus,
\begin{equation}
\label{eq:matched_pairs_sampling_dist}
\delta^{\new}_p \stackrel{(d)}{=} \epsilon^{\new}_p|\delta_p^\obs|,
\end{equation}
where the $\epsilon^{\new}_p$ are iid Rademacher random variables, independent of everything else. Let $P^\obs = \sum_{p=1}^{n/2} |\delta_p^\obs|$ be the observed number of nonzero pair differences. Then, by \eqref{eq:matched_pairs_sampling_dist},
\begin{equation}
\label{eq:reference_distribution_matched_pairs}
\set{n\hat \tau^{\new}/2 \mid P^\obs = r} \stackrel{(d)}{=} S_r,
\end{equation}
the random walk at time $r$. 

We now turn to understanding the law of $\hat \tau^\obs.$ Using casework, we can characterize the distribution of $\delta_p^\obs$ as a function of the 16 possible combinations of potential outcomes in pair $p, (Y_{p1}(1),Y_{p1}(0),Y_{p0}(1),Y_{p0}(0)).$ For example, suppose the table for pair $p$ has the form
\[
\begin{tabular}{c|cc}
        & $Y_i(1)$ & $Y_i(0)$ \\ \hline
$p1$   &    $a$      &    $b$      \\
$p0$   &    $b$      &    $a$ 
\end{tabular}
\]
for $a,b \in \set{0,1}.$ With probability $1/2$, the observed data is $(a,a)$, which forces $\delta_p^\obs = 0$. The same is true in the other case when $(b,b)$ is observed. When the table is of the form
\[
\begin{bmatrix}
a & 0 \\
1 & a
\end{bmatrix}, 
\quad
\begin{bmatrix}
1 & a \\
a & 0
\end{bmatrix},
\]
$\delta_p^\obs \stackrel{(d)}{=} B$, a fair Bernoulli random variable. When it is of the form
\[
\begin{bmatrix}
a & 1 \\
0 & a
\end{bmatrix}, 
\quad
\begin{bmatrix}
0 & a \\
a & 1
\end{bmatrix},
\]
$\delta_p^\obs \stackrel{(d)}{=} -B$. The last tables to handle are
\[
\begin{bmatrix}
1 & 1 \\
0 & 0
\end{bmatrix}, 
\quad
\begin{bmatrix}
0 & 0 \\
1 & 1
\end{bmatrix},
\quad\begin{bmatrix}
1 & 0 \\
1 & 0
\end{bmatrix}, 
\quad
\begin{bmatrix}
0 & 1 \\
0 & 1
\end{bmatrix}.
\]
In the first two cases, $\delta_p^\obs \stackrel{(d)}{=} \epsilon$. In the last two cases, $\delta_p^\obs = 1$ and $\delta_p^\obs = -1$ respectively. Table \ref{tab:matched-pair-distributions} summarizes the distributions of $\delta_p^\obs$ as a function of the potential outcomes of units $p1,p0$.

\begin{table}[ht]
\centering
\renewcommand{\arraystretch}{1.5}
\begin{tabular}{||c|cccc||}
\hline
\diagbox[dir=NW,width=7em,height=3em,linewidth=\arrayrulewidth]{Unit 1}{Unit 0}
& $(0,0)$ & $(0,1)$ & $(1,0)$ & $(1,1)$ \\
\hline
$(0,0)$
& $0$
& $-B$
& $B$
& $\epsilon$ \\

$(0,1)$
& $-B$
& $-1$
& $0$
& $-B$ \\

$(1,0)$
& $B$
& $0$
& $1$
& $B$ \\

$(1,1)$
& $\epsilon$
& $-B$
& $B$
& $0$ \\
\hline
\end{tabular}
\caption{Distributions of $\delta_p^\obs$ for all sixteen binary potential-outcome
tables of a matched pair $p$. Here, $B\sim\operatorname{Bernoulli}(1/2)$ and $\epsilon$ is
a Rademacher variable.}
\label{tab:matched-pair-distributions}
\end{table}

Define $\Delta_{pi} := Y_{pi}(1)-Y_{pi}(0)$. Inspecting Table \ref{tab:matched-pair-distributions}, the sum of the individual treatment effects for the units in the pair satisfies
\begin{equation}
\label{eq:matched_pairs_ITEs_and_D_p}
\Delta_{p0} + \Delta_{p1}
=
\begin{cases}
+2, & \text{if } \delta_p^\obs \stackrel{(d)}{=} +1,\\
+1, & \text{if } \delta_p^\obs \stackrel{(d)}{=} B,\\
0,  & \text{if } \delta_p^\obs \stackrel{(d)}{=} 0
      \text{ or } \delta_p^\obs \stackrel{(d)}{=} \epsilon,\\
-1, & \text{if } \delta_p^\obs \stackrel{(d)}{=} -B,\\
-2, & \text{if } \delta_p^\obs \stackrel{(d)}{=} -1.
\end{cases}
\end{equation}

We will construct a partition $\cl{G}$ of the pairs $p = 1,\dots,n/2$ into $K$ disjoint sets $G_k$ for some $K$, as follows. Let $\cl{I}_{+B}$ be the set of all pairs for which $\delta_p^\obs$ has the same distribution as $B$. Define the same for $\cl{I}_{-B}, \cl{I}_{-1}, \cl{I}_{+1}, \cl{I}_{0}, \cl{I}_{\epsilon}$. Let $m_{+B},m_{-B},m_{-1},m_{+1},m_{\epsilon}$ be the sizes of these sets. Under $H_{0N},$ \eqref{eq:matched_pairs_ITEs_and_D_p} shows
\begin{equation}
\label{eq:weak_null_forces_relation_on_pair_types}
    m_{+B} - m_{-B} + 2(m_{+1} - m_{-1}) = 0.
\end{equation}
Suppose $m_{+1} - m_{-1}  = \nu \geq 0.$ For as many pairs $p \in \cl{I}_{+1}$ as possible, match them with other pairs $p' \in \cl{I}_{-1}$. For such a pair $\set{p,p'}$, let this be another set in $\cl{G}$. The number of leftover pairs in $\cl{I}_{+1}$ is $\nu$. For every pair $p \in \cl{I}_{+B}$, match it with a pair $p' \in \cl{I}_{-B}$ and add $\set{p,p'}$ to the partition $\cl{G}.$ By \eqref{eq:weak_null_forces_relation_on_pair_types}, there are $2\nu$ unmatched pairs $p$ left over in $\cl{I}_{-B}$. For every $2$ leftover pairs $p_1,p_2$ in $\cl{I}_{-B}$, group them with an unmatched pair $p_3 \in \cl{I}_{+1}$ and add $\set{p_1,p_2,p_3}$ to the partition $\cl{G}$. Finally, for all other pairs in $\cl{I}_{\epsilon},\cl{I}_0$, add them as singleton sets into $\cl{G}$. For every set $G \in \cl{G}$, the sum $\sum_{p \in G} \delta^\obs_p$ is distributed as one of the following:
\begin{equation}
0, \quad \epsilon, \quad B - B', \quad 1 - B - B', \quad
\end{equation}
where $B,B'$ are independent Bernoulli$(1/2)$ random variables. Moreover, the sums are independent across groups $G$. Construct an analogous partition if $m_{+1} - m_{-1} < 0$. In this case, instead of $+1 - B - B'$ being a possible law, the option is $B + B' - 1$. Crucially, in all of these cases, the distribution is supported on $\set{-1,0,1}$ and symmetric around zero: for each $G \in \cl{G}$,
\begin{equation}
\sum_{p \in G} \delta^\obs_p \stackrel{(d)}{=} \left|\sum_{p \in G} \delta^\obs_p \right|\epsilon,
\end{equation}
for an independent Rademacher random variable $\epsilon.$ \sloppy Therefore, conditional on $P_{\cl{G}}^\obs := \sum_{G \in \cl{G}} \left|\sum_{p \in G} \delta^\obs_p \right|$,
\begin{equation}
\label{eq:matched_pairs_DIM_sampling distribution}
    n\hat \tau^\obs/2 = \sum_{G \in \cl{G}} \sum_{p \in G} \delta_p^\obs \stackrel{(d)}{=} S_{P_{\cl{G}}^\obs}.
\end{equation}

Putting our analyses of the sampling distribution and reference distribution together, Lemma \ref{lemma:random_walk_tail_comparison} shows that for every $t > 0,$
\begin{equation}
\Prob_{A^\new}(\hat \tau(A^\new,Y(A^\obs))\geq t) \geq \Prob(\hat \tau^\obs \geq t \mid P^\obs_{\cl{G}}).
\end{equation}
Indeed, $P_{\cl{G}}^\obs \equiv \sum_{G \in \cl{G}} \sum_{p \in G} \delta^\obs_p \equiv \sum_{p=1}^{n/2} \delta_p^\obs \equiv P^\obs \mod 2$ using the fact that $x \equiv |x| \mod 2$. By the triangle inequality, $P_{\cl{G}}^\obs \leq P^\obs.$ This shows \eqref{eq:tail_dominance_criterion} with $C = P^\obs_{\cl{G}}$; applying Proposition \ref{prop:tail_dominance} completes the proof.

\end{proof}

\subsection{Proofs for Unbalanced Experiments}

\begin{proof}[Proof of Theorem \ref{thm:unbalanced_experiments}]
Let $\rho$ be the treatment probability of the design. Under $H_{\leq 0N}$, there exists a pairing between all $n_{10}$ units with potential outcomes $(Y_i(1),Y_i(0)) = (1,0)$ and a subset of the $n_{01}$ units with outcomes $(0,1)$. Let $\pi$ be the permutation which swaps the units in each pairing and leaves all other units untouched. Define for any vectors $a,b \in \set{0,1}^n$ the difference
\begin{equation}
\begin{split}
D(a,b) & = n\rho(1-\rho)\left[\hat \tau(\pi a,Y(b)) - \hat \tau(b,Y(b)) \right] 
        = \sum_{i=1}^n (a_{\pi(i)} - b_i)\left(Y_i(b_i) - \frac{1}{2}\right).
\end{split}
\end{equation}
Then the exchangeability of $A^\new$ gives
\begin{equation}
p_F^+(A^\obs) = \Prob_{A^{\new}}( D(A^\new, A^\obs) \geq 0).
\end{equation}

Following the coverage proof of the Jackknife+ \cite{barber2021predictive}, interpret $D(a,b), D(b,a)$ as the scores in a two-way match between vectors $a,b \in \set{0,1}^n$ in a tournament played between all vectors in $\set{0,1}^n$. Declare $a$ to be a winner in the match if $D(a,b) \geq 0$, and $b$ to be a winner if $D(b,a) \geq 0$.
Observe that there can, in principle, be zero, one, or two winners.
Let $\cl{S}$ be the set of vectors whose opponent wins with probability at most $\alpha$ under $\scr{A}$:
\[
\cl{S} := \set{a \in \set{0,1}^n: \Prob_{b \in \scr{A}}(D(b,a) \geq 0) \leq \alpha }.
\]
Then $\Prob(p_F^+(A^\obs) \leq \alpha)$ is equal to $\Prob(A^\obs \in \cl{S})$. We claim that there always exists at least one winner in a match, via the bound
\begin{equation}
\label{eq:at_least_one_winner_per_match}
D(a,b) + D(b,a) \geq 0.
\end{equation}

Consider a pair $i,j$ where $\pi(i) = j, \pi(j) = i$ and $(Y_i(1),Y_i(0)) = (1,0), (Y_j(1),Y_j(0)) = (0,1)$. Then the contribution of indices $i,j$ to $D(a,b) + D(b,a)$ is equal to
\begin{align*}
    & (a_{j} - b_i)\left(Y_i(b_i) - \frac{1}{2}\right) + (a_{i} - b_j)\left(Y_j(b_j) - \frac{1}{2}\right) + \\
    & (b_{j} - a_i)\left(Y_i(a_i) - \frac{1}{2}\right) + (b_{i} - a_j)\left(Y_j(a_j) - \frac{1}{2}\right) \\
   = &  (a_{j} - b_i)\left(Y_i(b_i) - Y_j(a_j)\right) + (a_i - b_j) \left(Y_j(b_j) - Y_i(a_i)\right) = 0.
\end{align*}
The last equality follows because $Y_i(t)=t$ and $Y_j(t)=1-t$ for $t\in\{0,1\}$, so both terms vanish by the identity $(x-y)(x+y-1)=0$ for all $x,y\in\{0,1\}$.

For any single unit $i$, the contribution of unit $i$ to 
$D(a,b) + D(b,a)$ is $(a_i - b_i)(Y_i(b_i) - Y_i(a_i))$. When $Y_i(1) = Y_i(0)$, the contribution is zero; when $(Y_i(1),Y_i(0)) = (0,1)$, the contribution is $\mathbf{1}\set{a_i \neq b_i}$. This proves \eqref{eq:at_least_one_winner_per_match}. Finally, 
\begin{align*}
    \Prob(A^\obs \in \cl{S})^2 & = \Prob(A^\obs \in \cl{S}, A^\new \in \cl{S})\\
    & \leq \E\left[\mathbf{1}\set{D(A^\obs,A^\new) \geq 0} + \mathbf{1}\set{D(A^\new,A^\obs) \geq 0}; A^\obs \in \cl{S}, A^\new \in \cl{S}  \right] \\
    & \leq 2\E\left[\mathbf{1}\set{D(A^\new,A^\obs) \geq 0}; A^\obs \in \cl{S}, A^\new \in \cl{S}  \right] \\
    & \leq 2\E\left[\mathbf{1}\set{D(A^\new,A^\obs) \geq 0}; A^\obs \in 
    \cl{S}  \right] 
     \leq 2\E\left[p_F^+; A^\obs \in \cl{S}  \right] 
     \leq 2\alpha \Prob(A^\obs \in \cl{S}).
\end{align*}
If $\Prob(A^\obs \in \cl{S}) = 0$, the claim is immediate. Otherwise, dividing by $\Prob(A^\obs \in \cl{S})$ yields the desired bound.

\end{proof}

\subsection{Anticonservativeness in the Bounded Case}

\begin{example}
\label{ex:anti_conservative}

For each design considered in this paper, we give examples with nonbinary potential outcomes in which the level-$\alpha$ test based on $p_F^+$ is anticonservative. Similar constructions show that $p_F^\pm$ can also be anticonservative.

For the CRE, take $n = 6$ with $(Y_i(1))_{i=1}^6 = (0,0,1/2,1/2,1/2,1/2)$ and $(Y_i(0))_{i=1}^6 = (0,0,0,0,1,1)$, so $\tau = 0.$ Consider $p_F^+$ with $\alpha = 0.05$. Let $E$ be the event where the treated set is either $\set{3,5,6}$ or $\set{4,5,6}$. On $E$, the observed data consists of $(1/2,1/2,1/2)$ in the treated arm and $(0,0,0)$ in the control arm. In the FRT reference distribution, the observed difference-in-means value of $1/2$ is the unique largest value obtainable upon randomization. The one-sided FRT always rejects on this event and $\Prob(E) = 0.10,$ violating the level $\alpha$.

For the BRE, take $n = 6$ and consider $4$ units with type $(0,0)$ and two units with types $(3/4,1/4), (1/4,3/4).$ By \eqref{eq:alternative_representation_cHT}, the sampling distribution of $\hat \tau_{\cHT}$ is $S_4/6$, where $S_r$ is the simple random walk at time $r$. For every observed assignment, \eqref{eq:alternative_representation_cHT} again shows that the reference distribution of the FRT is $\left(S_4'+ \e/2 + \e'/2\right)/6$, with $\e,\e'$ independent Rademacher random variables. On the event $S_4 = 4$, which happens with probability $1/16$, $p_F^+ = \Prob(S_4'+ \e/2 + \e'/2 \geq 4) = 3/64 < 0.05$. Therefore the one-sided FRT rejects with probability at least $1/16$.

For matched pairs, take $n=12$ units in $6$ pairs. In each of the first four pairs, let the two units have types $(1,1)$ and $(0,0)$. In the fifth pair, both units have type $(1/2,0)$, and in the sixth, both have type $(0,1/2)$. Then $\tau=0$, and the observed treated-minus-control differences in the last two pairs are always $1/2$ and $-1/2$. Consequently, the sampling distribution of $n\hat\tau_{\DIM}/2$ is $S_4$, whereas its FRT reference distribution is $S_4'+\e/2+\e'/2$. The same calculation as for the BRE gives $p_F^+ = 3/64 < 0.05$ on an event of probability $1/16$.
\end{example}

\end{document}